\documentclass[conference]{IEEEtran}
\IEEEoverridecommandlockouts
\usepackage{tabularx}

\usepackage{cite}
\usepackage{amsmath,amssymb,amsfonts}
\usepackage[hyphens]{url}
\usepackage{algorithmic}

\usepackage{textcomp}
\usepackage{amsthm}
\theoremstyle{definition}
\newtheorem{definition}{Definition}[section]
\usepackage{subcaption}
\usepackage{xcolor}
\usepackage{caption}
\usepackage{graphicx}

\def\BibTeX{{\rm B\kern-.05em{\sc i\kern-.025em b}\kern-.08em
    T\kern-.1667em\lower.7ex\hbox{E}\kern-.125emX}}
    \makeatletter
\let\old@ps@IEEEtitlepagestyle\ps@IEEEtitlepagestyle
\def\confheader#1{%
    \def\ps@IEEEtitlepagestyle{%
        \old@ps@IEEEtitlepagestyle%
        \def\@oddhead{\strut\hfill#1\hfill\strut}%
        \def\@evenhead{\strut\hfill#1\hfill\strut}%
    }%
    \ps@headings%
}
\makeatother

\confheader{%
    \parbox{20cm}{Accepted by 2023 IEEE International Workshop on Computer Aided Modeling and Design of Communication Links and Networks (CAMAD), \textcopyright 2023 IEEE}
}
\begin{document}

\setlength{\tabcolsep}{5pt}
\renewcommand{\arraystretch}{1.5}
\title{How to synchronize Digital Twins? A Communication Performance Analysis}


\author{
	\IEEEauthorblockN{Lal Verda Cakir\IEEEauthorrefmark{1}\IEEEauthorrefmark{2},Sarah Al-Shareeda\IEEEauthorrefmark{1},Sema F. Oktug\IEEEauthorrefmark{1},Mehmet Özdem\IEEEauthorrefmark{3},Matthew Broadbent\IEEEauthorrefmark{4},and Berk Canberk\IEEEauthorrefmark{4}}
	\IEEEauthorblockA{\IEEEauthorrefmark{1}Department of Computer Engineering, Istanbul Technical University, Istanbul, Turkey
	}
 \IEEEauthorblockA{\IEEEauthorrefmark{2}BTS Group, Istanbul, Turkey
	}
    \IEEEauthorblockA{\IEEEauthorrefmark{3} Türk Telekom,  Turkey
	}
    \IEEEauthorblockA{\IEEEauthorrefmark{4} School of Computing, Engineering and The Built Environment, Edinburgh Napier University, UK
	}
	Email: \{cakirl18, alshareeda, oktug\}@itu.edu.tr, verda.cakir@btsgrp.com, mehmet.ozdem@turktelekom.com.tr, \\  \{M.Broadbent, B.Canberk\}@napier.ac.uk

}
\maketitle
\begin{abstract}
Synchronization is fundamental for mirroring real-world entities in real-time and supporting effective operations of Digital Twins (DTs). Such synchronization is enabled by the communication between the physical and virtual realms, and it is mostly assumed to occur in real-time. However, this is not the case, as real-life scenarios witness performance degradation that may lead to synchronization problems. Hence, as such a problem has yet to be thoroughly analyzed in the literature, this work attempts to uncover potential challenges by emulating and analyzing the DT traffic flows in networks of different scales, for different communication protocols, and with various flow configurations. We propose a Twin Alignment Ratio metric to evaluate the synchronization performance to achieve this goal. Consequently, the findings reveal the interplay of network infrastructure, protocol selection, and twinning rate on synchronization and performance.
\end{abstract}

\begin{IEEEkeywords}
Digital Twins, Synchronization, Twinning Rate, Communication, Performance Analysis, Simulations
\end{IEEEkeywords}

\section{Introduction}\label{intro}
Achieving synchronization is crucial to ensure that the Digital Twins (DTs) accurately mirror real-world entities in real-time, forming the foundation for reliable decision-making and effective operational network management \cite{dt_survey}. The current literature has assumed a real-time synchronization to be present. However, in the pursuit of implementing DTs in real life, the actual synchronization cannot be achieved due to the bottleneck caused by communication delays and inaccuracies; DT communication requires enormous network flows operating simultaneously and continuously, and hence, divergent traffic patterns from traditional network traffic emerge. In this sense, an evaluation methodology is required to quantify such traffic change and to assess performance. Network performance analysis has been emphasized to address the impact of varying protocols and flow configurations to uncover insights that can guide the development of tailored communication strategies. Additionally, the assessment of alignment between physical and virtual entities is required to identify the synchronization issues within DTs.

Based on this motivation, this paper aims to emulate and analyze the traffic flows in the network that are generated to accommodate the integration of the DT concept. This can be accomplished via the traditional delay, jitter, and packet loss metrics. Nonetheless, an extra specialized metric is required to precisely identify the potential challenges of implementing DTs. In this sense, our main contributions to the literature can be listed as:
\begin{itemize}
    \item We implement a simulation framework for DT communication and conduct analysis for various network topologies, protocols, flow configurations, and end-user network occupancy,
    \item We analyze the performance of DT communication in terms of the traditional delay, jitter, and packet loss metrics and
    \item We introduce a new Twin Alignment Ratio $\tau$ metric to precisely assess the DT communications synchronization performance and identify future challenges of DT communication.
\end{itemize}

The remainder of this paper is organized such that Section \ref{literature} reviews the literature on network performance analysis with the presence of DT communication. Section \ref{lab:tech} describes the simulation system model. Section \ref{analysis} showcases and discusses the simulation results. Finally, Section \ref{conc} concludes the paper and provides future directions.
\section{Related Work}\label{literature}
This literature review explores recent advancements encompassing modeling and emulating network traffic, optimizing network performance, and predictive insights for proactive traffic flow estimation.

Emulating and modeling traffic flows, Yang et al. \cite{yang2021systematic} lay the foundation with a flow emulation framework tailored for DT networks. This framework addresses the precise emulation of network traffic with the expanding bandwidth and speed in physical networks, ensuring synchronization between the physical and DT network traffic. Moreover, Ak et al. \cite{ak2023t6conf} introduce a DT networking framework based on IPv6 infrastructure for synchronization in DT-based smart city applications. Almeida et al. \cite{nuno2022machine} integrate a machine learning propagation loss module for wireless network DTs, aiding in creating authentic DT environments for evaluating networking solutions. Leveraging the potential of AI, Ferriol-Galmes et al. \cite{ferriol2022flowdt} introduces a deep-learning solution that models network flows predicting per-flow performance metrics to enhance accuracy in network traffic pattern analyses. Similarly, Li et al.'s approach \cite{li2023learnable} streamlines network configuration estimation using learning-based techniques, efficiently mapping network configurations to performance indicators.

Regarding performance optimization, Tang et al. \cite{tang2022intelligent} delve into the complexities of achieving low latency and deterministic bandwidth within DT networks. Their method handles the scheduling of delay-sensitive network traffic, alleviating congestion and enhancing the conveyance of high-priority traffic. Similar considerations in \cite{van2022fairness} address fairness-aware edge computing and latency minimization. Their novel optimization approach minimizes latency in the context of ultra-reliable and low-latency communications, optimizing bandwidth allocation, transmit power, task offloading policies, and processing rates for equitable resource allocation. Progressing into AI-driven optimization, Ferriol-Galmes et al. \cite{ferriol2022building} introduce TwinNet. This Graph Neural Network-based model excels in discerning complex relationships among network elements, making substantial contributions to Quality of Service (QoS) optimization. To optimize the network performance by reducing interference, Cakir et al. \cite{dtwn} propose a transmit power control methodology based on DT wireless networks, considerably increasing the throughput. Moreover, Duran et al. \cite{topology_discovery} offered a DT-enriched green discovery policy incorporating energy consumption predictions to optimize the topology discovery in core networks. In this proposed methodology, the energy consumption and number of visited ports are sustainably decreased.

Exploring predictive solutions for traffic estimation, Ono et al. \cite{ono2023amond} proposes a networking scheme that harmonizes mobile ad-hoc networks and DTs, reducing congestion while preserving communication speed and stability. The pursuit of efficient computation offloading leads Van Huynh et al. \cite{van2022urllc} to design an iterative algorithmic approach rooted in convex approximation, optimizing resource allocation in DT wireless edge networks. Moreover, Masaracchia et al. \cite{dt_6g} consider DTs as an enabler for resource allocation methodologies thanks to their precise decision-making abilities. Similarly, Sun et al. \cite{resource_allocation} utilizes DTs for resource allocation in vehicular networks. Moreover, Saravanan et al. \cite{saravanan2022enabling} use neural networks to predict per-path mean delay in communication networks, resulting in a scalable network DT that enhances overall network performance. Lai et al. \cite{lai2023deep} leverage deep learning for network traffic prediction, improving the accuracy of background traffic matrices and optimizing synchronization within DT networks.
\section{Proposed Technical Scheme}\label{lab:tech}
The DT literature defines the synchronization frequency between the physical device and DT as the twinning rate \cite{twinning_rate}. It is generally considered a parameter that is adjusted to achieve high-accuracy twins. It can be implemented on physical devices as such; however, various challenges may arise, which cause the potential inability to maintain this desired twinning rate of the twin. 

The DT communication occurs as the data packets from twinned devices traverse the network to reach the device's destination in the DT's location. In this process, the propagation delays, routing overhead, and queuing at intermediate network nodes contribute to the end-to-end delay of the packet and cause variability of this delay. Also, while trying to meet the real-time requirements of DT, network congestion can cause packet drops in network nodes and increase delay and jitter. Moreover, the prominence of these challenges may change depending on the communication protocol used, the twinning rate implemented, and the network topology. All of these factors will contribute to a level of synchronization discrepancies, leading to DTs' performance degradation.

\subsection{Twin Alignment Ratio}
Considering these aspects and challenges, constructing DT systems based on predefined twinning rates is short-sighted, and evaluating the twinning capability based on achieved twinning rates needs to be revised. Therefore, we incorporate the concept of "attempts vs. achieved" to provide a more comprehensive assessment of synchronization. Here, to allow a deeper understanding of the communication reliability in DT communication, we define the Twin Alignment Ratio $\tau$ based on the following:

\begin{definition}[Planned Twinning Frequency, $f_p$]
    The frequency at which the twined device (e.g. IoT device) is configured to schedule the transmission of the packet per second.
\end{definition}

\begin{definition}[Achieved Twinning Frequency, $f_a$]
    The frequency at which the DT received the packets sent from the twinned device per second. This frequency is regarded as the actual synchronization outcome with the assumption that the processing delay of the update at DT will be uniformly distributed in time.
\end{definition}

Consequently, we define the Twin Alignment Ratio $\tau$ as:
\begin{definition}[Twin Alignment Ratio, $\tau$]
    The ratio of Achieved Twinning Frequency $f_a$ to Planned Twinning Frequency $f_p$.
    \begin{equation}
        \tau = \frac{f_a}{f_p}
    \end{equation}
\end{definition}
\begin{figure*}
    \centering
    \includegraphics[width=\linewidth]{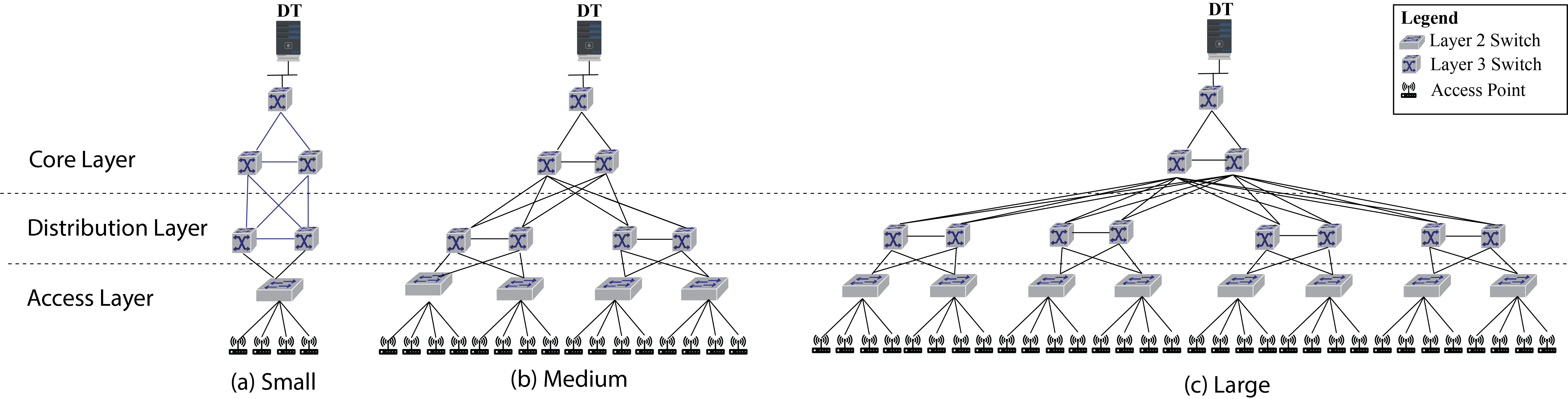}
    \caption{Simulated Network Topologies}
    \label{fig:topology}
\end{figure*}
\subsection{Network Typologies}
\label{lab:network_topo}
The network scale is highly correlated with the performance of individual data flows and the Twin Alignment Ratio $tau$ due to increased simultaneous communication links. Therefore, we implement three topologies, namely, small, middle, and large, as depicted in Fig. \ref{fig:topology}, which are designed using the topology structure defined in \cite{cisco_network} with the architectural components:
\begin{itemize}
    \item \textbf{Core Layer:} The backbone of the network is created by two Layer 3 switches and performs the routing and switching operations of the Distribution Layer.
    \item \textbf{Distribution Layer:} The management of the access layer is performed by enforcing policies and doing access control. Here, there are a minimum of two Layer 3 switches and a point-to-point link connecting each pair.
    \item \textbf{Access Layer:} This layer consists of Layer 2 switches and Access Points (APs) connected to these switches. Access to the network is given by this layer to the devices. This layer provides end-users and IoT devices a point of entry to the network and manages network resources and policies. 
\end{itemize}

The Wireless Local Area Networks (WLANs) enable the twinned devices' connectivity, including other end-users. These WLANs are configured to operate in distinctive channels with station mobility within the coverage of their first associated AP. Moreover, the stations (twinned devices and end-users) are uniformly distributed among and within WLANs. The thresholds where performance might degrade can be pinpointed by dissecting these topologies at different scales. Furthermore, the outcomes of this evaluation can guide decisions related to resource allocation, traffic management, and system resilience, all of which are vital components of robust DT communication networks.

\subsection{Addressed Communication Protocols}
DT communication is required to have real-time responsiveness while ensuring reliable data delivery. These two distinctive traits are present in the User Datagram Protocol (UDP) and Transmission Control Protocol (TCP) protocols. UDP is a connectionless protocol with low overhead and might provide minimal latency thanks to its lightweight nature. On the other hand, TCP, as a connection-oriented protocol, guarantees reliable data delivery by establishing a virtual connection and employing retransmission mechanisms. This is accomplished by TCP's three-way handshake mechanism using synchronization and acknowledgment processes, which can introduce higher latency and jitter. As a result, TCP might impact real-time responsiveness but provide more reliability than UDP. Due to this one-to-one mapping of these protocols' requirements and features, a tradeoff can occur when other parameters are included. Therefore, we evaluate UDP and TCP performance with applications set up on the Twinned Devices (\textit{TD}) with packet generation scheduled according to the Planned Twinning Frequency $f_p$.
Moreover, we set the End-user Devices (\textit{ED}) to communicate at randomized times with the DT. Here, the link is formed by an application over TCP, where the end users are the receivers. This application setup reflects the scenario of end users utilizing the services of DT.

\section{Simulation-based Comparative Analysis}\label{analysis}
This study delves into an in-depth assessment of the performance of DT communication, focusing on the QoS for communication flows and the Twin Alignment Ratio under varying Planned Twinning Frequencies (\textit{$f_p$}) and topology scales. To conduct these investigations, simulations were carried out using ns-3 \cite{Ns3}, a discrete-event network simulator, with the parameters specified in Table \ref{tab:sim}. The simulations encompassed both UDP and TCP flows for the \textit{TD} scenario, considering instances with (w/) and without (w/o) background traffic (\textit{BT}) generated by \textit{UD} across the three distinct topologies outlined in Section \ref{lab:tech}.

\begin{table}[!htbp]
    \begin{center}\caption{Simulation Parameters}    \label{tab:sim}
    \begin{tabular}{|l|l|}
    \hline
       \textbf{Parameter}  & \textbf{Value} \\
       \hline
       Simulation Duration & 20 sec \\
       Number of APs & \{4, 16, 32\}\\
       Stations (STA) density & 10 STAs per AP\\
       \textit{TD} / \textit{UD}& 0.5\\
       Channel Band & 2.4Ghz \\
       Mobility & As described in Section \ref{lab:network_topo}\\
       Payload & 1024 bytes \\
       Data Rate of L3 Link & 1Mbps \\
       Delay of L3 Link & 2 msec \\
       Data Rate of L2 Link & 500Kbps \\
       Delay of L2 Link & 1 msec \\
       $f_p$  & \{0.5 pps,1 pps,2 pps, 5 pps,10 pps\}\\
       \textit{UD} Application On Time  &  Uniform Random Variable [0, 2]\\
       \textit{UD} Application Off Time  &  Uniform Random Variable [0, 0.5]\\
       \hline
    \end{tabular}
    \end{center}
\end{table}

\begin{figure}[!htbp]
    \begin{subfigure}[b]{0.5\linewidth}
        \centering
        \includegraphics[width=\linewidth]{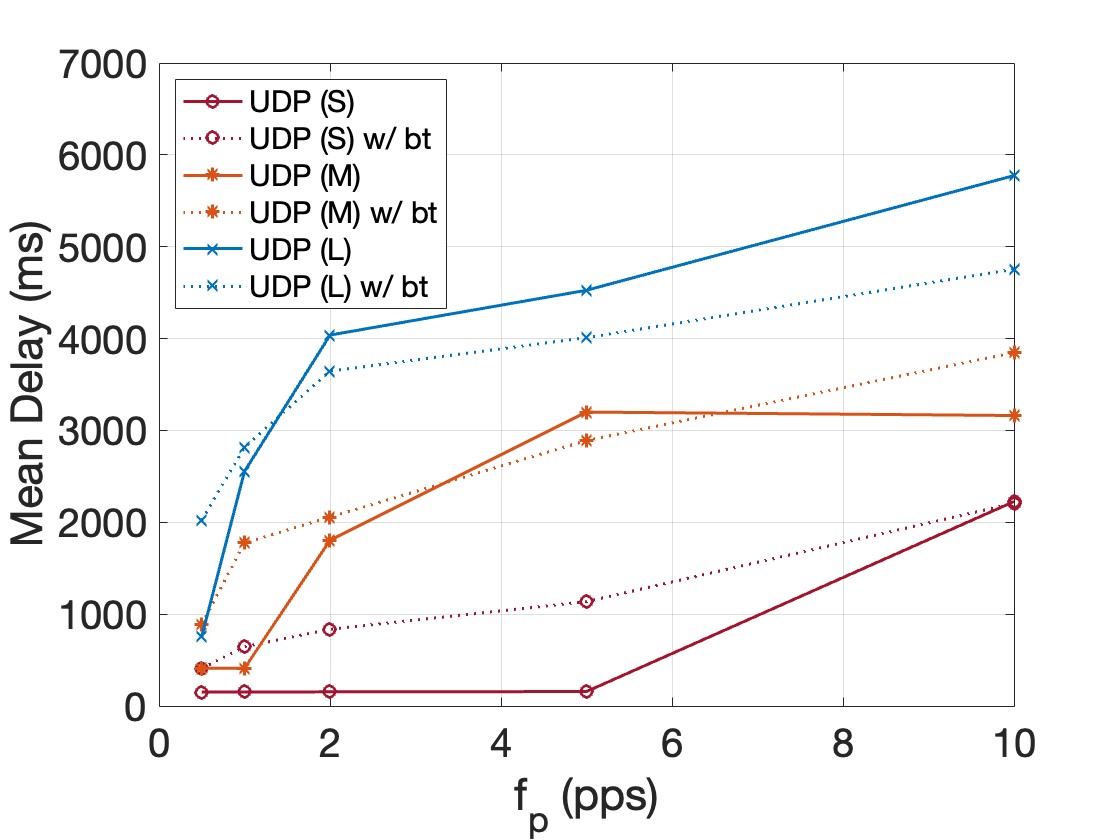}
        \caption{Delay - UDP}\label{fig:udp_delay}
    \end{subfigure}%
    \begin{subfigure}[b]{0.5\linewidth}
        \centering
        \includegraphics[width=\linewidth]{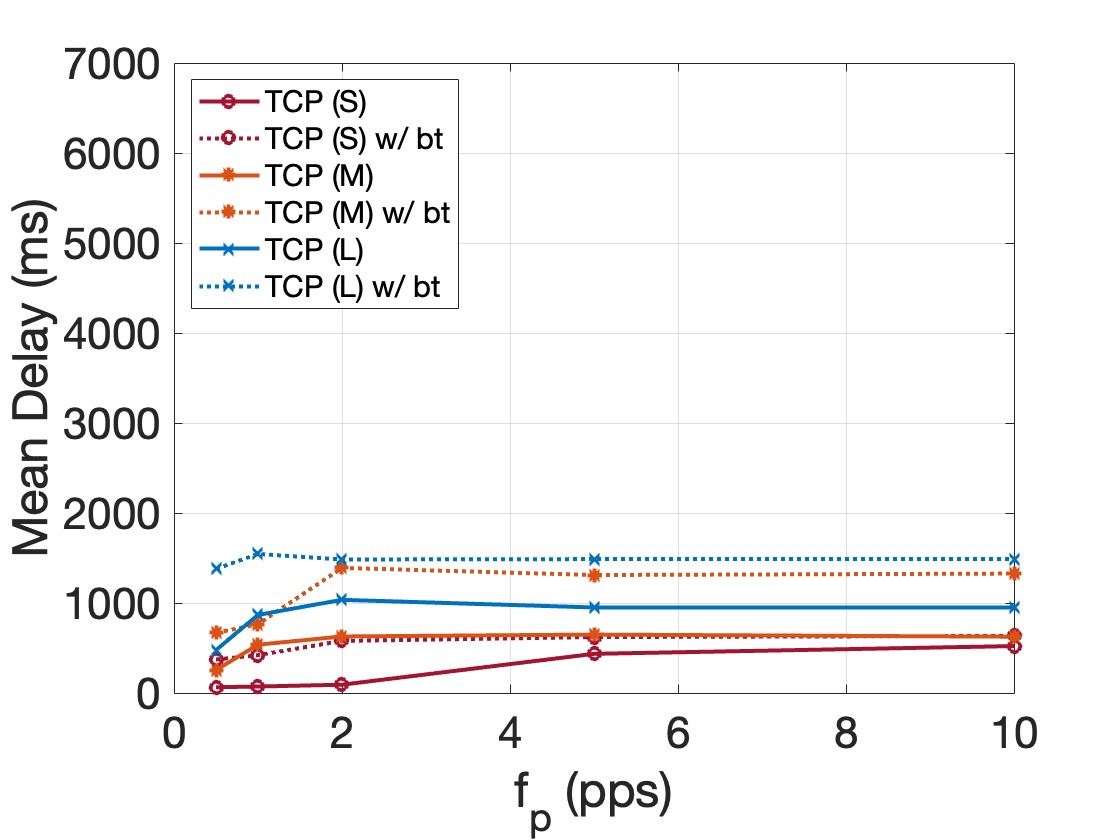}
        \caption{Delay - TCP}\label{fig:tcp_delay}
    \end{subfigure}
    \hfill
    \begin{subfigure}[b]{0.5\linewidth}
        \centering
        \includegraphics[width=\textwidth]{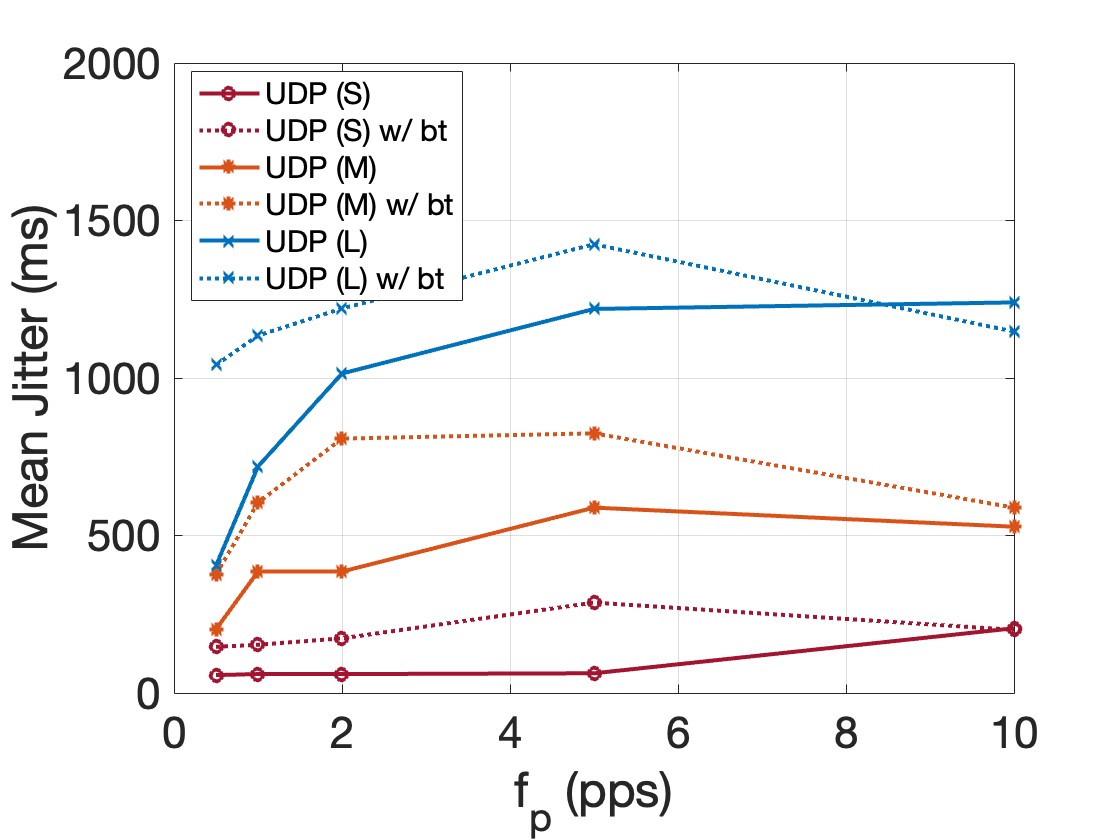}
        \caption{Jitter - UDP}\label{fig:udp_jitter}
    \end{subfigure}%
    \begin{subfigure}[b]{0.5\linewidth}
        \centering
        \includegraphics[width=\linewidth]{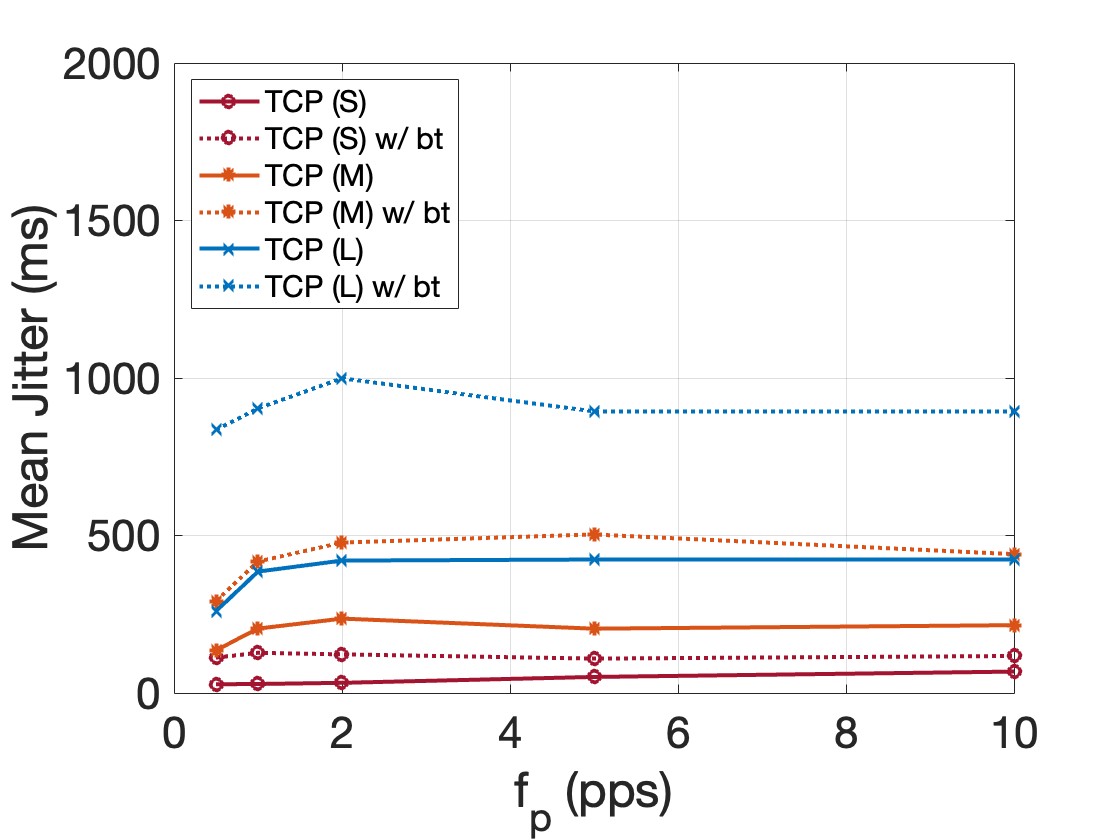}
        \caption{Jitter - TCP}\label{fig:tcp_jitter}
    \end{subfigure}%
    \caption{Delay and Jitter Comparison}
    \label{fig:delay_jitter}
\end{figure}

In terms of the delay and jitter performance metrics outlined in Section \ref{lab:tech}, the simulation results in Fig. \ref{fig:udp_delay} and Fig. \ref{fig:udp_jitter} illustrate an increase in both indicators for UDP flows as the topology scale expands. This phenomenon is attributed to the heightened network utilization caused by $f_p$ and concurrent Background Traffic (\textit{BT}), leading to higher queuing delays at intermediate nodes and higher WLAN contention. In contrast, TCP flows exhibit consistent and relatively low mean delay and jitter under various conditions, as demonstrated in Fig. \ref{fig:tcp_delay} and Fig. \ref{fig:tcp_jitter}. This outcome is potentially attributed to TCP's intrinsic flow and congestion control mechanisms that regulate data rates, ensuring stable performance. Nevertheless, this regulation may hinder the full execution of the set $f_p$.

\begin{figure}[!htbp]
    \begin{subfigure}[b]{0.5\linewidth}
        \centering
        \includegraphics[width=\linewidth]{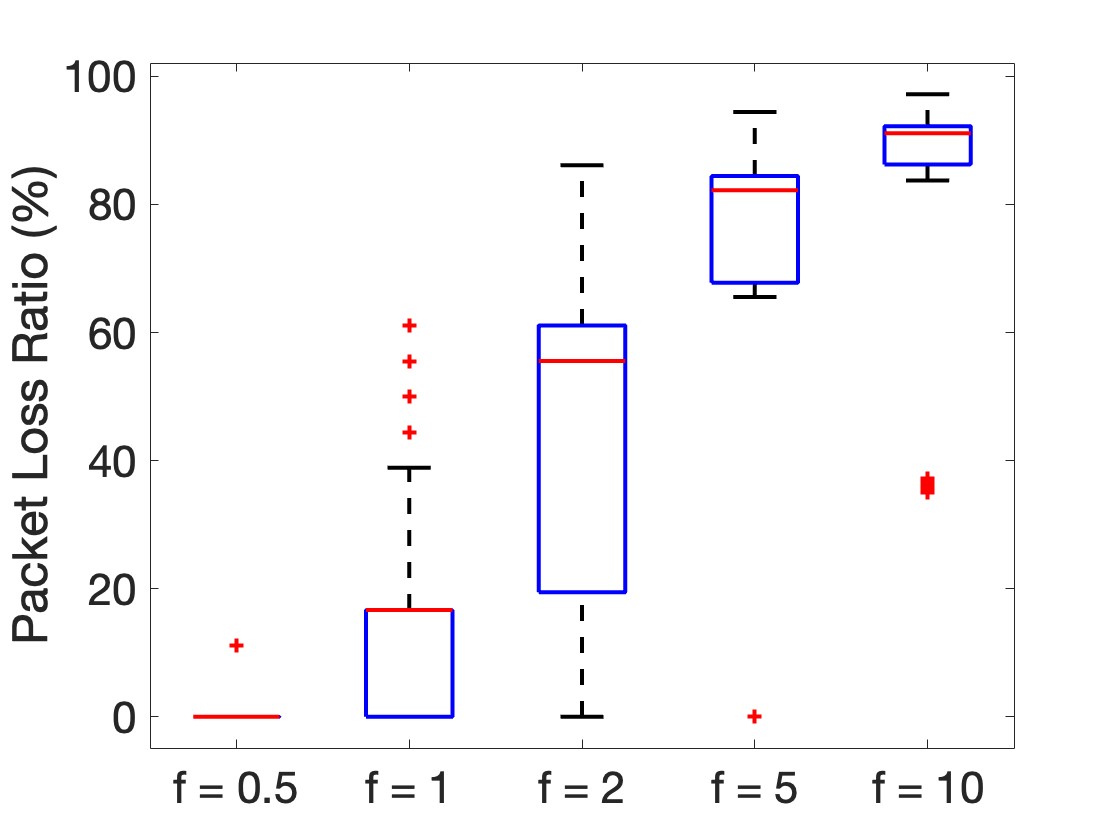}
        \caption{UDP}\label{fig:udp_delay1}
    \end{subfigure}%
    \begin{subfigure}[b]{0.5\linewidth}
        \centering
        \includegraphics[width=\linewidth]{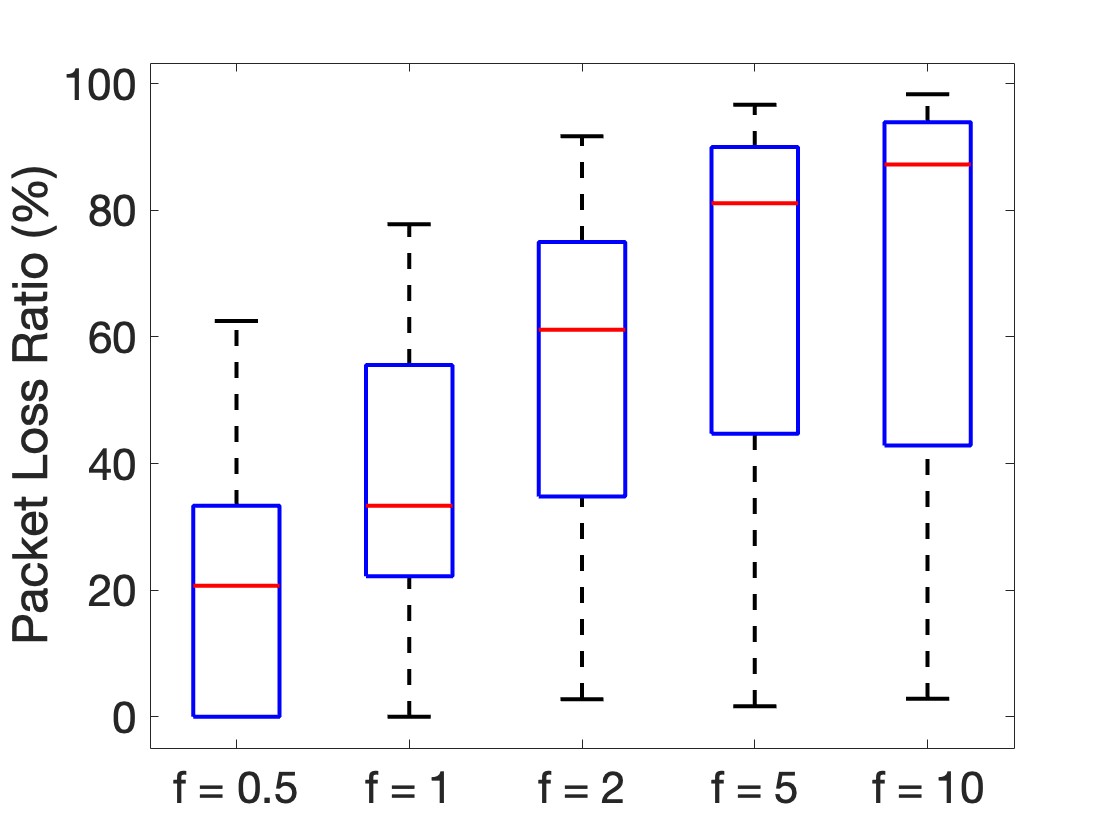}
        \caption{UDP w/ bt}\label{fig:tcp_delay1}
    \end{subfigure}
    \\
    \begin{subfigure}[b]{0.5\linewidth}
        \centering
        \includegraphics[width=\textwidth]{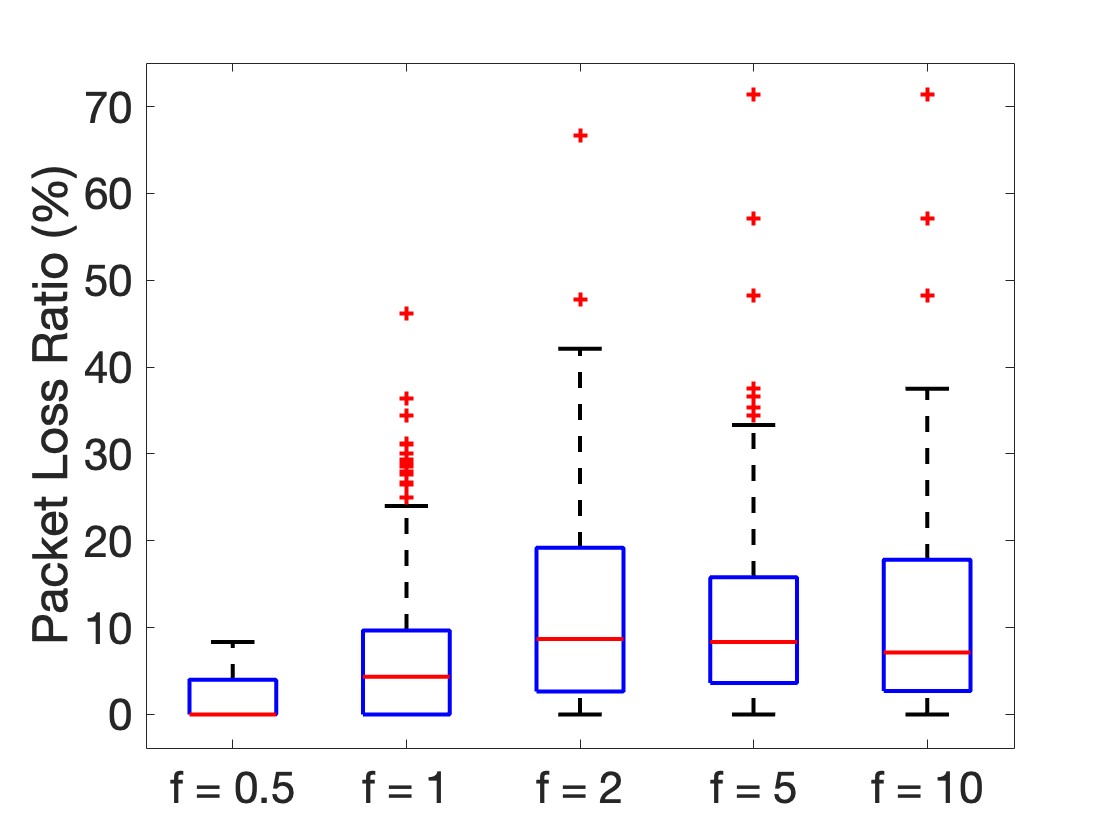}
        \caption{TCP}\label{fig:tcp_box}
    \end{subfigure}%
    \begin{subfigure}[b]{0.5\linewidth}
        \centering
        \includegraphics[width=\linewidth]{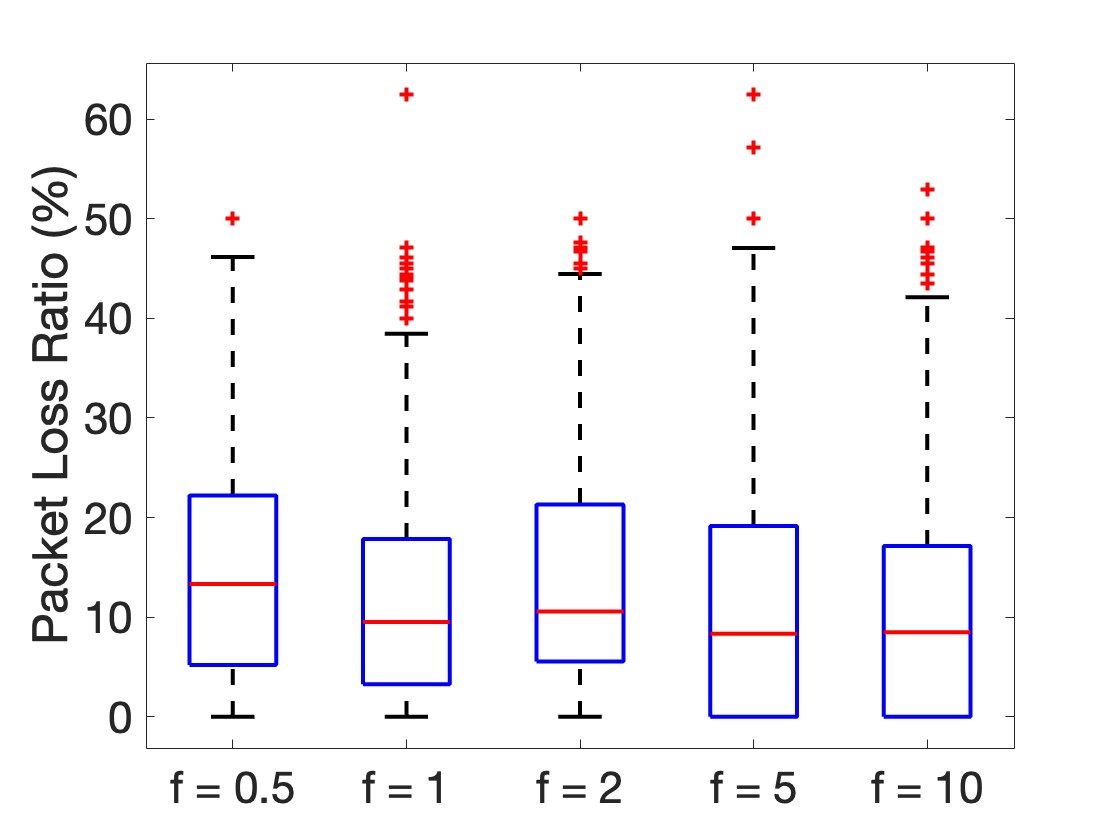}
        \caption{TCP w/ bt}\label{fig:tcp_box_b}
    \end{subfigure}%
    \caption{Packet Loss Comparison}
    \label{fig:packet_loss}
\end{figure}

\begin{figure}[!htbp]

    \begin{subfigure}[b]{0.5\linewidth}
        \centering
        \includegraphics[width=\linewidth]{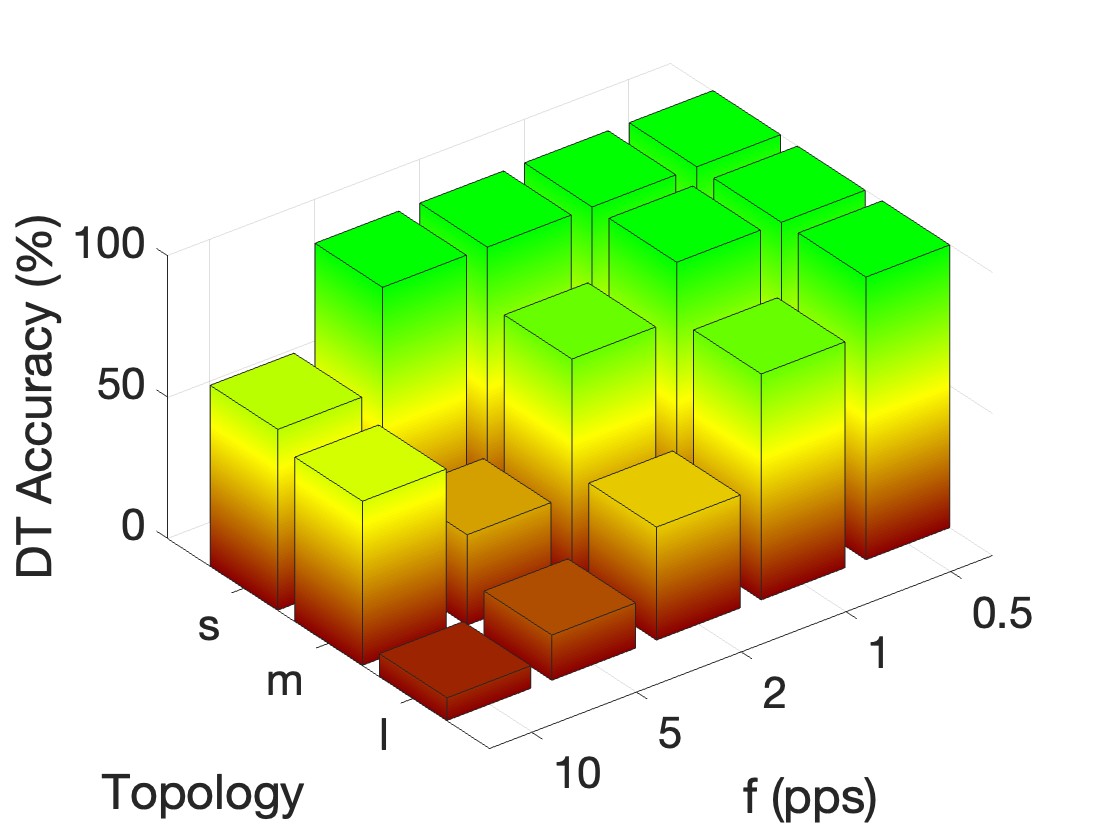}
        \caption{UDP w/o bt}\label{fig:udp_dt}
    \end{subfigure}%
    \begin{subfigure}[b]{0.5\linewidth}
        \centering
        \includegraphics[width=\linewidth]{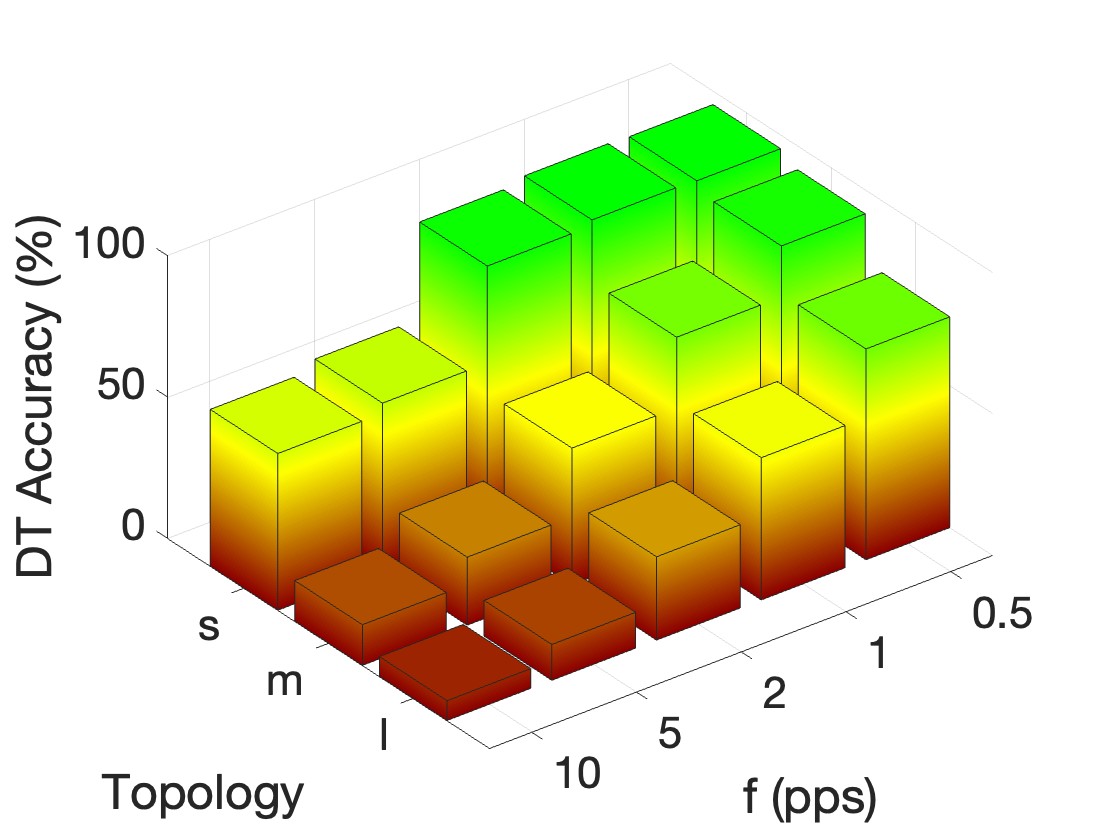}
        \caption{UDP w/ bt}\label{fig:udp_dt_b}
    \end{subfigure}%
        \\
    \begin{subfigure}[b]{0.5\linewidth}
        \centering
        \includegraphics[width=\linewidth]{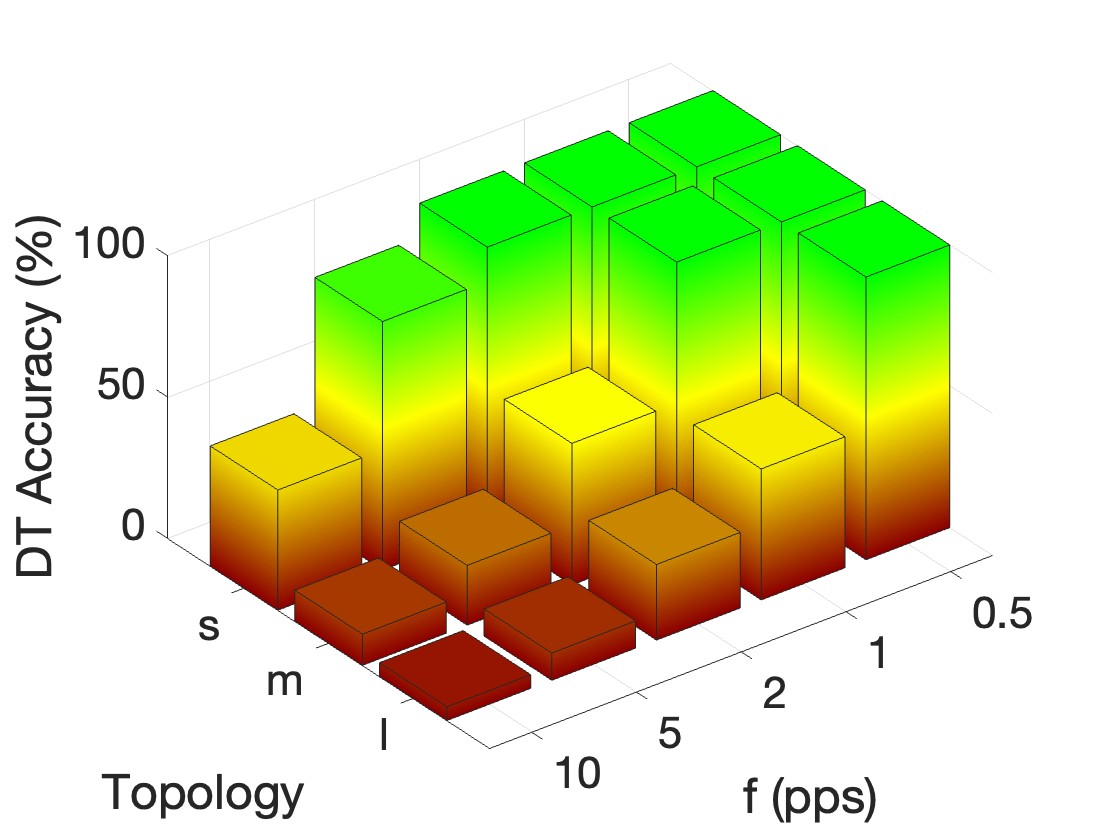}
        \caption{TCP w/o bt}\label{fig:tcp_dt}
    \end{subfigure}%
    \begin{subfigure}[b]{0.5\linewidth}
        \centering
        \includegraphics[width=\linewidth]{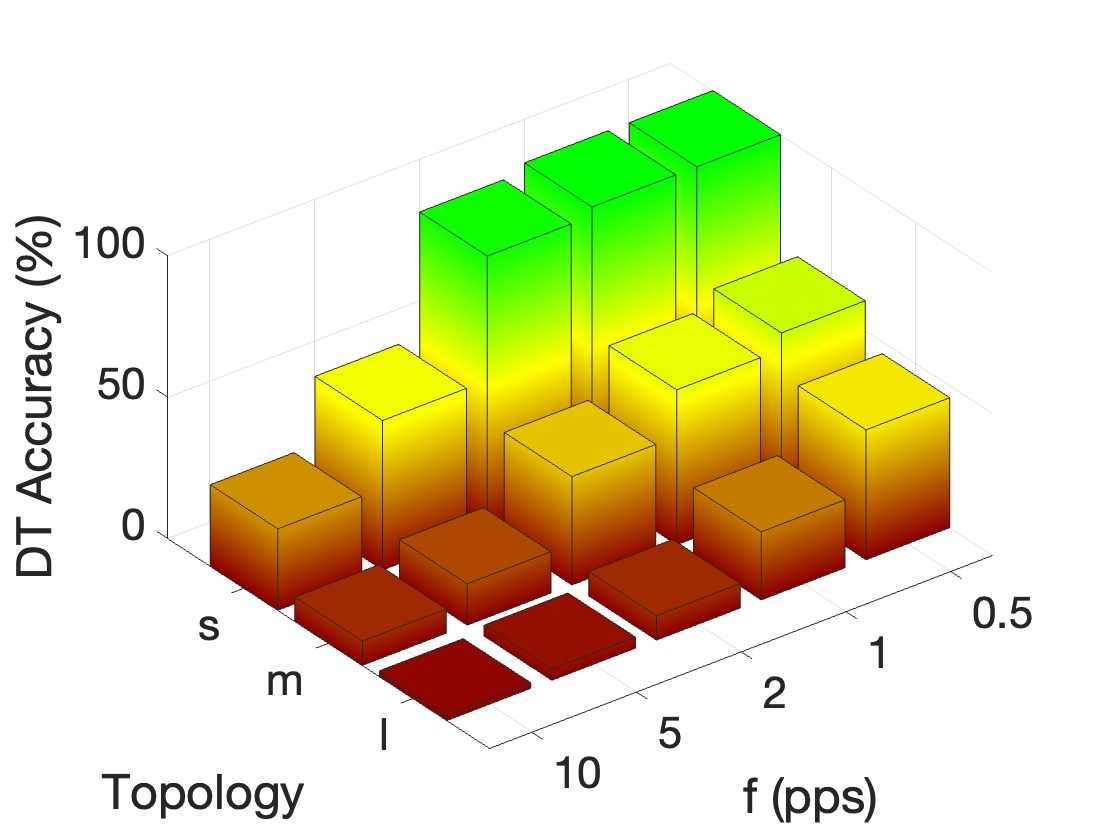}
        \caption{TCP w/ bt}\label{fig:tcp_dt_b}
    \end{subfigure}%
    \caption{Twin Alignment Ratio ($\tau$) Performance Metric}
    \label{fig:twinalignment}
\end{figure}

Considering the packet loss ratio distribution among flows, the UDP performance is notably affected by the level of \textit{BT}, packet loss in wireless communication, and increased packet drops in the network, as can be seen in Fig. \ref{fig:packet_loss}. In contrast, the TCP's proficiency in retransmission and acknowledgment mechanisms supports a lower uniformly distributed packet-loss ratio. One point to hint at is that the TCP protocol's packet loss statistics exhibit a presence of more outliers, which is further exacerbated by adding the \textit{BT}. Such a case can be attributed to the TCP's complex congestion control and retransmission mechanisms, which can lead to occasional delays or retransmissions in the presence of \textit{BT} or changing network dynamics. Furthermore, it could be because of randomized traffic generation from the \textit{UD} devices; this is not true with UDP's packet loss distribution.

Lastly, Fig. \ref{fig:twinalignment} compares the Twin Alignment Ratio between TCP and UDP. Across both protocols, the $\tau$ values remain consistent as $f_p$ decreases, and the network scale diminishes. Conversely, there is a noticeable decline in $\tau$ with the escalation of $f_p$, topology scale, and \textit{BT}. Moreover, for TCP, an additional reduction in $\tau$ occurs due to the protocol's connection-oriented nature, which restricts the transmission of essential packets during simulation, thereby contributing to the decline in $\tau$. It is worth highlighting that the observed trend in UDP's $\tau$ closely aligns with the insights from the packet loss analysis, a characteristic attributed to the transmission of data packets.


As a summary of these results, When examining the impact of DT communication, these findings indicate that TCP is preferable over UDP, especially when consistent performance and fewer disruptions in QoS metrics are desired. On the other hand, UDP provides a best-effort twinning process, which is preferable when the network utilization is high.

\section{Conclusion and Future Extension}\label{conc} 
In this study, we investigated the performance of DT communication, examining aspects such as individual flow delay, jitter, packet loss, and synchronization. Notably, we introduced a novel metric called the Twin Alignment Ratio to quantitatively assess synchronization performance in DTs, addressing a research gap. We uncovered significant insights by comparing TCP and UDP flows across various network sizes and considering background traffic. TCP demonstrated consistent performance and better QoS, while UDP exhibited pronounced twin alignment. Surprisingly, attempts to enhance real-time performance through increased twinning frequency reduced synchronization performance. This study underscores the crucial impact of network infrastructure, communication protocol selection, and twinning frequency on QoS and synchronization outcomes. These findings offer valuable guidance for optimizing DT communication performance. In terms of future directions, exploring adaptive twinning strategies and advanced synchronization techniques could further enhance the overall effectiveness of DT communication.

\section*{Acknowledgment}
This work was supported by DeepMind Scholarship Program, 100/2000 YÖK Ph.D. Scholarship Program, ITU Rektorlugu Bilimsel Arastirma Projeleri Birimi (BAP) Fund under Grant Number 43981, and The Scientific and Technological Research Council of Turkey (TUBITAK) 1515 Frontier R\&D Laboratories Support Program for BTS Advanced AI Hub: BTS Autonomous Networks and Data Innovation Lab. Project 5239903

\bibliographystyle{IEEEtran}
\bibliography{bibfile.bib}

\begin{thebibliography}{10}
\providecommand{\url}[1]{#1}
\csname url@samestyle\endcsname
\providecommand{\newblock}{\relax}
\providecommand{\bibinfo}[2]{#2}
\providecommand{\BIBentrySTDinterwordspacing}{\spaceskip=0pt\relax}
\providecommand{\BIBentryALTinterwordstretchfactor}{4}
\providecommand{\BIBentryALTinterwordspacing}{\spaceskip=\fontdimen2\font plus
\BIBentryALTinterwordstretchfactor\fontdimen3\font minus
  \fontdimen4\font\relax}
\providecommand{\BIBforeignlanguage}[2]{{%
\expandafter\ifx\csname l@#1\endcsname\relax
\typeout{** WARNING: IEEEtran.bst: No hyphenation pattern has been}%
\typeout{** loaded for the language `#1'. Using the pattern for}%
\typeout{** the default language instead.}%
\else
\language=\csname l@#1\endcsname
\fi
#2}}
\providecommand{\BIBdecl}{\relax}
\BIBdecl

\bibitem{dt_survey}
B.~R. Barricelli, E.~Casiraghi, and D.~Fogli, ``A survey on digital twin:
  Definitions, characteristics, applications, and design implications,''
  \emph{IEEE Access}, vol.~7, pp. 167\,653--167\,671, 2019.

\bibitem{yang2021systematic}
H.~Yang, Y.~Li, K.~Yao, T.~Sun, and C.~Zhou, ``A systematic network traffic
  emulation framework for digital twin network,'' in \emph{2021 IEEE 1st
  International Conference on Digital Twins and Parallel Intelligence
  (DTPI)}.\hskip 1em plus 0.5em minus 0.4em\relax IEEE, 2021, pp. 94--97.

\bibitem{ak2023t6conf}
E.~Ak, K.~Duran, O.~A. Dobre, T.~Q. Duong, and B.~Canberk, ``T6conf: Digital
  twin networking framework for ipv6-enabled net-zero smart cities,''
  \emph{IEEE Communications Magazine}, vol.~61, no.~3, pp. 36--42, 2023.

\bibitem{nuno2022machine}
E.~Nuno~Almeida, M.~Rushad, S.~R. Kota, A.~Nambiar, H.~L. Harti, C.~Gupta,
  D.~Waseem, G.~Santos, H.~Fontes, R.~Campos \emph{et~al.}, ``Machine learning
  based propagation loss module for enabling digital twins of wireless networks
  in ns-3,'' \emph{arXiv e-prints}, pp. arXiv--2205, 2022.

\bibitem{ferriol2022flowdt}
M.~Ferriol-Galm{\'e}s, X.~Cheng, X.~Shi, S.~Xiao, P.~Barlet-Ros, and
  A.~Cabellos-Aparicio, ``Flowdt: A flow-aware digital twin for computer
  networks,'' in \emph{ICASSP 2022-2022 IEEE International Conference on
  Acoustics, Speech and Signal Processing (ICASSP)}.\hskip 1em plus 0.5em minus
  0.4em\relax IEEE, 2022, pp. 8907--8911.

\bibitem{li2023learnable}
B.~Li, T.~Efimov, A.~Kumar, J.~Cortes, G.~Verma, A.~Swami, and S.~Segarra,
  ``Learnable digital twin for efficient wireless network evaluation,''
  \emph{arXiv preprint arXiv:2306.06574}, 2023.

\bibitem{tang2022intelligent}
Z.~Tang, D.~Chen, T.~Sun, L.~Zhang, M.~Qi, and X.~Wang, ``Intelligent awareness
  of delay-sensitive internet traffic in digital twin network,'' \emph{IEEE
  Journal of Radio Frequency Identification}, vol.~6, pp. 891--895, 2022.

\bibitem{van2022fairness}
D.~Van~Huynh, V.-D. Nguyen, S.~R. Khosravirad, and T.~Q. Duong,
  ``Fairness-aware latency minimisation in digital twin-aided edge computing
  with ultra-reliable and low-latency communications: A distributed
  optimisation approach,'' in \emph{2022 56th Asilomar Conference on Signals,
  Systems, and Computers}.\hskip 1em plus 0.5em minus 0.4em\relax IEEE, 2022,
  pp. 1045--1049.

\bibitem{ferriol2022building}
M.~Ferriol-Galm{\'e}s, J.~Su{\'a}rez-Varela, J.~Pailliss{\'e}, X.~Shi, S.~Xiao,
  X.~Cheng, P.~Barlet-Ros, and A.~Cabellos-Aparicio, ``Building a digital twin
  for network optimization using graph neural networks,'' \emph{Computer
  Networks}, vol. 217, p. 109329, 2022.

\bibitem{dtwn}
\BIBentryALTinterwordspacing
L.~V. Çakır, K.~Huseynov, E.~Ak, and B.~Canberk, ``Dtwn: Q-learning-based
  transmit power control for digital twin wifi networks,'' \emph{EAI Endorsed
  Transactions on Industrial Networks and Intelligent Systems}, vol.~9, no.~31,
  p.~e5, Jun. 2022. [Online]. Available:
  \url{https://publications.eai.eu/index.php/inis/article/view/1059}
\BIBentrySTDinterwordspacing

\bibitem{topology_discovery}
K.~Duran and B.~Canberk, ``Digital twin enriched green topology discovery for
  next generation core networks,'' \emph{IEEE Transactions on Green
  Communications and Networking}, pp. 1--1, 2023.

\bibitem{ono2023amond}
S.~Ono, T.~Yamazaki, T.~Miyoshi, A.~Taya, Y.~Nishiyama, and K.~Sezaki, ``Amond:
  Area-controlled mobile ad-hoc networking with digital twin,'' \emph{IEEE
  Access}, 2023.

\bibitem{van2022urllc}
D.~Van~Huynh, V.-D. Nguyen, S.~R. Khosravirad, V.~Sharma, O.~A. Dobre, H.~Shin,
  and T.~Q. Duong, ``Urllc edge networks with joint optimal user association,
  task offloading and resource allocation: A digital twin approach,''
  \emph{IEEE Transactions on Communications}, vol.~70, no.~11, pp. 7669--7682,
  2022.

\bibitem{dt_6g}
A.~Masaracchia, V.~Sharma, B.~Canberk, O.~A. Dobre, and T.~Q. Duong, ``Digital
  twin for 6g: Taxonomy, research challenges, and the road ahead,'' \emph{IEEE
  Open Journal of the Communications Society}, vol.~3, pp. 2137--2150, 2022.

\bibitem{resource_allocation}
W.~Sun, P.~Wang, N.~Xu, G.~Wang, and Y.~Zhang, ``Dynamic digital twin and
  distributed incentives for resource allocation in aerial-assisted internet of
  vehicles,'' \emph{IEEE Internet of Things Journal}, vol.~9, no.~8, pp.
  5839--5852, 2022.

\bibitem{saravanan2022enabling}
M.~Saravanan, P.~S. Kumar, and A.~R. Kumar, ``Enabling network digital twin to
  improve qos performance in communication networks,'' in \emph{2022 IEEE
  Smartworld, Ubiquitous Intelligence \& Computing, Scalable Computing \&
  Communications, Digital Twin, Privacy Computing, Metaverse, Autonomous \&
  Trusted Vehicles (SmartWorld/UIC/ScalCom/DigitalTwin/PriComp/Meta)}.\hskip
  1em plus 0.5em minus 0.4em\relax IEEE, 2022, pp. 2151--2160.

\bibitem{lai2023deep}
J.~Lai, Z.~Chen, J.~Zhu, W.~Ma, L.~Gan, S.~Xie, and G.~Li, ``Deep learning
  based traffic prediction method for digital twin network,'' \emph{Cognitive
  Computation}, pp. 1--19, 2023.

\bibitem{twinning_rate}
\BIBentryALTinterwordspacing
D.~Jones, C.~Snider, A.~Nassehi, J.~Yon, and B.~Hicks, ``Characterising the
  digital twin: A systematic literature review,'' \emph{CIRP Journal of
  Manufacturing Science and Technology}, vol.~29, pp. 36--52, 2020. [Online].
  Available:
  \url{https://www.sciencedirect.com/science/article/pii/S1755581720300110}
\BIBentrySTDinterwordspacing

\bibitem{cisco_network}
\BIBentryALTinterwordspacing
May 2021. [Online]. Available:
  \url{https://www.cisco.com/c/en/us/td/docs/solutions/Enterprise/Campus/HA\_campus\_DG/hacampusdg.html}
\BIBentrySTDinterwordspacing

\bibitem{Ns3}
\BIBentryALTinterwordspacing
 [Online]. Available: \url{https://www.nsnam.org/}
\BIBentrySTDinterwordspacing

\end{thebibliography}
\end{document}